\documentclass[useAMS,usenatbib]{mn2e}
\usepackage{graphicx}
\usepackage{txfonts}

\title{WR 143: A Wolf-Rayet Binary}
\author[W. P. Varricatt \& N. M. Ashok]
       {Watson P. Varricatt$^{1}$\thanks{E-mail: w.varricatt@jach.hawaii.edu(WPV);
               ashok@prl.ernet.in(NMA)}
               \& Nagarhalli M. Ashok$^{2}$\footnotemark[1]  \\
        $^{1}$Joint Astronomy Centre, 660 N. Aohoku Place, Hilo,  Hawaii 96720, USA\\
        $^{2}$Physical Research Laboratory, Navrangpura, Ahmedabad, India 380009}
\date{Accepted 2005 October.
      Received 2005 ????;
            in original form 2005 June 20}

\pagerange{\pageref{firstpage}--\pageref{lastpage}}
\pubyear{2005}

\begin{document}

\maketitle

\label{firstpage}

\begin{abstract}
Near infrared spectroscopy and photometry of the Wolf-Rayet Star WR 143 
(HD 195177) were obtained in the $JHK$ photometric bands.  High resolution 
spectra observed in the $J$ and $H$ bands exhibit narrow 1.083-$\mu$m  
He\,{\sc i} line and the H\,{\sc i} Pa$\beta$ and the Brackett series 
lines in emission superposed  on the broad emission line spectrum of
the Wolf-Rayet star, giving strong indications of the presence  of a
companion.  From the narrow emission lines observed, the companion 
is identified to be an early-type Be star. The photometric magnitudes
exhibit variations in the $JHK$ bands which are probably due to the 
variability of the companion star. The flux density distribution is 
too steep for a Wolf-Rayet atmosphere.   This is identified  to be 
mainly due to the increasing contribution from the early-type companion
star towards shorter wavelengths.
\end{abstract}


\newpage

\section{Introduction}

During the late stages of evolution, massive stars 
(with M $\geq$ 30 M$_{\odot}$) go through the Wolf-Rayet (WR) phase.  At this 
stage, they are subjected to large scale mass loss (\.{M} $\sim 10^{-5}$ 
M$_{\odot}$ year$^{-1}$) through accelerated stellar winds with terminal 
velocities in the range 750-5000 km s$^{-1}$. Consequently these objects are 
characterized by infrared excess and strong, broad  emission lines originating 
in their fast winds.  Based on the emission lines observed, they are classified
into WN, WC and WO stars. WN stars show lines of He and N in their winds with 
the evidence of H in the late types and WC and WO stars show lines of He, C and 
O. WR 143 (HD 195177) is an interesting member of the WC4 type, its lines being 
weaker than the rest of the members of its class.

Early studies classified WR 143 as WC5+(OB) (Smith 1968).  The possibility 
of the presence of a companion star was considered since its emission lines
were weaker than that of many other stars of the same WR type, even though 
no absorption lines were detected.  Smith, Shara \& Moffat (1990a)  
reclassified WR 143 as a WC4 star. Figer, McLean \& Najarro (1997) observed
the $K$-band spectrum. They also noticed that the IR emission lines of
WR 143 were weaker and broader compared to those of the other WC5-type stars,
WR 111 and WR 114.  The VIIth catalogue of galactic Wolf-Rayet Stars
(van der Hucht 2001) lists  WR 143 as WC4 + OB?.  Considering the 
absolute {$\upsilon$} magnitude of the system, van der Hucht (2001)
proposed a B0V companion to the WR star in WR 143. The exact nature of 
the companion is not yet understood. The object is very faint at radio 
and X-ray wavelengths. The 6-cm radio continuum survey of 
Abbott et al. (1986) using the VLA gave an upper limit of the flux of
0.4\,mJy, which was above their 3$\sigma$ noise.  However, the 3.6-cm 
radio continuum survey of Cappa, Goss and van der Hucht (2004) using
the VLA did not detect WR 143.  They derived an upper limit for its mass
loss rate to be 0.7 $\times$ 10$^{-5}$ M$_{\odot}$ year$^{-1}$.
WR 143 may have been detected (2$\sigma$) in the ROSAT X-ray survey 
(Pollock, Haberl, Corcoran 1995).  

WR 143 is located close to the galactic plane at a distance of -1\,pc 
(van der Hucht et al. 1988). The distance estimates by different investigators
agree quite well.  From the spectroscopic parallax, Conti \& Vacca (1990) 
estimated a heliocentric distance of 1.0\,kpc which is close to the value
of 0.82\,kpc estimated by van der Hucht et al. (1988) and 1.17\,kpc by
Smith, Shara \& Moffat (1990b) using the line flux method. 
van der Hucht et al. (1988) estimated the extinction A$_{\upsilon}$ = 6.07,
which gives A$_{V}$=5.47 assuming A$_{\upsilon}$/A$_{V}$=1.11 
(van der Hucht 2001).

\section{Observations and data reduction}
We observed the near-IR $JHK$ spectra of WR 143 with the 3.8-m United Kingdom 
Infrared Telescope (UKIRT), and the Cooled Grating Spectrometer (CGS4)
(Mountain et al. 1990) using the 40\,l/mm grating. This grating, with a 
1-pixel slit, gives a resolution of $\sim$940 in the $J$ band in the
second order and $\sim$680 and $\sim$900 in the $H$ and $K$ bands respectively 
in the first order.  The $J$-band spectra were taken on three epochs.  
Krypton, Xenon and Argon arc lamps' spectra were used to wavelength-calibrate
the $J$-, $H$- and $K$-band spectra respectively. 
Table {\ref {obslog}}. gives the details of the spectroscopic observations.  
Fig. {\ref{wr143j} shows the observed $J$-band spectra and Fig. \ref{wr143hk}, 
the $H$- and $K$-band spectra.   Many narrow emission features were observed 
in our $JHK$ spectra. To better understand these, we again observed at 
higher spectral resolution at these wavelengths.

Our $J$-band spectra on all three epochs exhibited a prominent narrow emission 
component superposed on the broad He\,{\sc i} emission line from the WR star
at 1.083\,$\mu$m. To investigate this feature, we carried out high-resolution
observations with CGS4 using the echelle grating and a 2-pixel slit, at a 
spectral resolution of $\sim$20700. Six wavelength settings with the echelle
grating gave a reasonably good coverage of the 1.083-$\mu$m He\,{\sc i} line,
although a part of the red wing could not be covered.  The observed
spectrum is shown in  Fig.{\ref{1431083.ps}.  The wavelength calibration was
carried out using the photospheric absorption lines of the comparison star.
Heliocentric corrections were applied.  The narrow emission component 
has an asymmetric profile and its central wavelength is seen
very close to the line-centre of the broad emission 
line from the WR star.  The low resolution $J$-band spectra 
taken over the three epochs with a period of 5 months do not 
show any noticeable shift in the location of the narrow emission component.
Visual inspection shows that this emission was also present at nearly the 
same wavelength in the spectrum presented by Eenens \& Williams (1994).
There is a conspicuous narrow absorption seen at 10778 $\AA$ which is not
identified. From the well defined blue edge of the P Cyg
absorption profile of the line, we estimate a V$_{edge}$= 2845 $\pm$ 15 km/s.
This is close to the value of V$_{\infty}=$ 2750 km/s observed by 
Eenens \& Williams (1994).

The 1.281-$\mu$m He\,{\sc ii} line also exhibited a faint narrow emission 
component close to the line-centre.  Some of the lines from 
1.55\,$\mu$m to 1.681\,$\mu$m were also observed to be much narrower than 
the rest of the WR lines. Observations were again carried out using UKIRT
and UIST (Ramsay Howat et al. 2000) at higher spectral resolution (R$\sim$4000)
to better resolve the narrow emission lines. The UIST spectra were wavelength 
calibrated using an Argon arc lamp mounted inside the instrument. These spectra 
show the H\,{\sc i} Pa\,$\beta$ and the Br series lines 
superposed on the broad emission line spectrum of the WR star.
Fig. \ref{wr143JH_U} shows the observed spectra in the $JH$ bands.  

All the spectroscopic observations were carried out by nodding the telescope 
on two positions separated by $\sim$12 arcsec along the slit.  The flat-field
observations were obtained by exposing the arrays to black bodies mounted 
inside the instruments. Preliminary reduction of the data including co-adding 
the frames and the bias and the flat-field corrections were carried out using 
ORACDR, the pipeline reduction facility at UKIRT. The spectra were optimally 
extracted using the STARLINK software FIGARO.  
Comparison stars (listed in Table \ref{obslog}) were observed at all 
wavelength settings. The observed comparison star spectra were corrected for 
their photospheric temperatures by dividing by appropriate black bodies and 
their photospheric hydrogen absorption lines were interpolated across and
removed at the continuum level before ratioing the object spectra with them.
The final reduction, involving the calculations of the line fluxes
and the equivalent widths (EWs) were carried out using IRAF.

\begin{table}
\begin{centering}
\caption{Spectroscopic Observations using UKIRT}
\label{obslog}
\begin{tabular}{@{}lllll}
\hline
$\lambda$       &Grating/ &Resolution	&UT Date  		&Comparison 	        \\
($\mu$m)        &grism(*) &($\lambda/\Delta \lambda$) &(yymmdd.ddd) &star  		\\
\hline
1.083       &Echelle    &20700          &011029.221		&BS 8170         	\\
1.18        &40\,l/mm    &944           &010613.568 		&BS 7756,		\\
	    &		&		&			&BS 8788		\\
1.18        &40\,l/mm    &944           &010704.424 		&BS 7793  		\\
1.18        &40\,l/mm    &944           &011029.368		&BS 8170 		\\
1.70        &40\,l/mm    &680           &010614.553		&BS 7793 		\\
2.25        &40\,l/mm    &900           &010614.625 		&BS 7793  		\\
1.237	    &long\_J*   &4100           &030619.418 		&BS 7672  		\\
1.522	    &short\_H*  &3800           &030619.435 		&BS 7672  		\\
1.702	    &long\_H*	&4000		&030812.415 		&BS 7672   		\\
\hline
\multicolumn{5}{l}{* Observed using UIST.  The rest were observed using CGS4 }\\
\end{tabular}
\end{centering}
\end{table}

\begin{figure}
\centering
\includegraphics[width=8.0cm,clip]{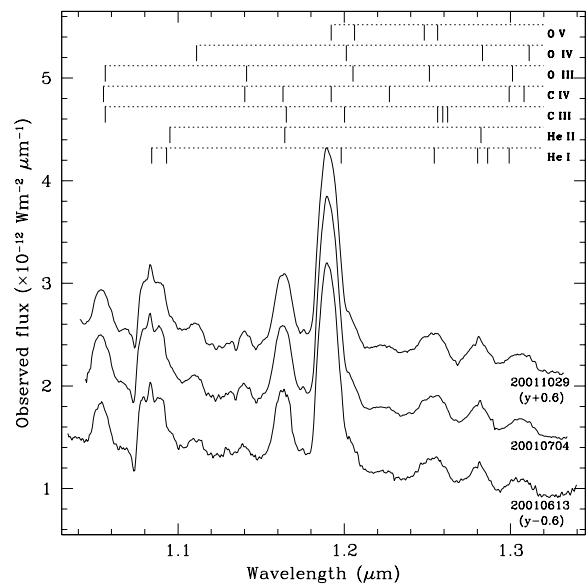}
\vskip -2cm
\caption{The observed $J$-band spectra of WR 143.  Spectra are labelled by the
UT-dates of the observations.  All spectra are plotted with the same scale, with those
of 20010613 and 20011029 vertically shifted for clarity.}
\label{wr143j}
\end{figure}

Photometry in the $JHK$ bands were acquired on 3 epochs from 2001 March
to 2002 December using the 1.2-m Mt. Abu Infrared Telescope and an 
LN2-cooled 256x256 NICMOS3 IR array.  FS 149 and FS 150, two of the UKIRT
faint standard stars (Hawarden et al. 2001), were observed for calibration.
Observations were carried out by dithering the object on several positions on 
the array. The dark observations were obtained before each set of the on-sky 
observations and the flat-field corrections were applied using the flat-fields
generated from the object observations by median-combining the 
observed frames. The observed $JHK$ magnitudes are shown in Table {\ref{mags}}
along with the 2MASS magnitudes and the other near-IR photometric 
measurements available.

The observed spectra were flux-calibrated using the average $JHK$ 
magnitudes of the three epochs of our observations. The strong emission 
lines present in the photometric bands contribute significantly to the 
observed magnitudes.  Hence, the magnitudes were corrected to subtract out 
the contribution from the emission lines adopting a method similar to 
Eenens and Williams (1992).  The equivalent widths (EW) of the emission 
lines were estimated from the ratioed spectra; these EWs were weighted with 
the transmission of the filters used in the Mt. Abu photometry and the 
corrections for the lines were derived as:

\noindent
$\Delta_{mag}$ = 2.5 log $ \frac{LW+FW}{FW}$

\noindent
where FW is the band width of the filter used for the photometry
and LW is the sum of the weighted EWs of the emission lines  within
the photometric band.  $\Delta_{mag}$ were added to the observed
magnitudes to obtain the magnitudes representing the continua, which
were then used to flux-calibrate the observed $JHK$ spectra.

\begin{table}
\caption{The near-IR photometric data}
\label{mags}
\begin{tabular}{@{}lllll}
\hline
Place and Date 		      &$J$	    &$H$		&$K$		&$L$    \\
\hline
Mt. Abu$^{\bullet}$         &8.18 (.06)   &7.61 (.03)         &6.99 (.05)     &	\\      
Mt. Abu$^{\diamond}$          &8.13 (.04)   &7.63 (.05)         &7.05 (.05)     &	\\      
Mt. Abu$^{\Delta}$            &8.15 (.04)  &7.65 (.04)   	&7.09 (.07)     &	\\
2MASS$^{\dagger}$  	      &8.59 (.01)  &8.10 (.01) 	&7.46 (.01)  	&	\\
Cohen et al. ('75)            &             &                   &7.50 (0.2)     &7.2 (0.2)	\\
Allen et al. ('72) 	      &		    &7.96		&7.34		&	\\
\hline
\multicolumn{5}{l}{$^{\bullet}$ 15 Dec., '02, $^{\diamond}$ 14 Dec., '02, $^{\Delta}$ 28 Mar.,'01, $^{\dagger}$ 21 June '98}\\
\end{tabular}
\end{table}

\begin{figure*}
\centering
\hspace*{-20pt}
\includegraphics[width=17.0cm,clip]{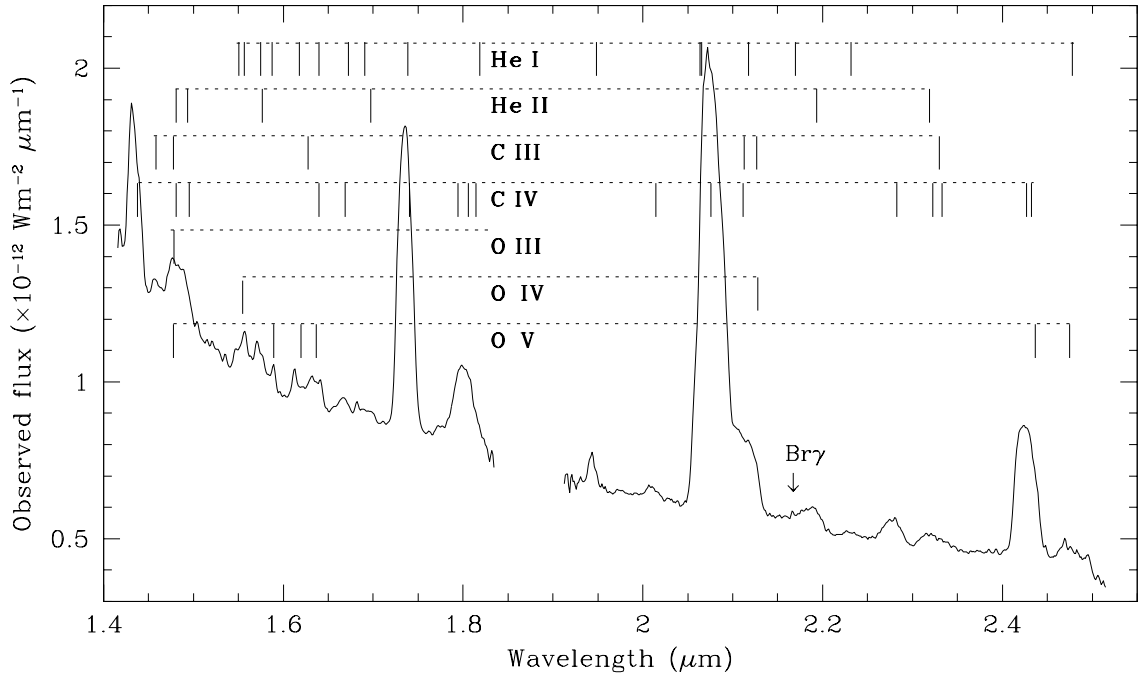}
\vspace*{-11cm}
\caption{The observed $H$- and $K$- band spectra of WR 143. The gap shows the
region of strong telluric absorption.}
\label{wr143hk}
\end{figure*}

\begin{figure}
\centering
\includegraphics[width=8.0cm,clip]{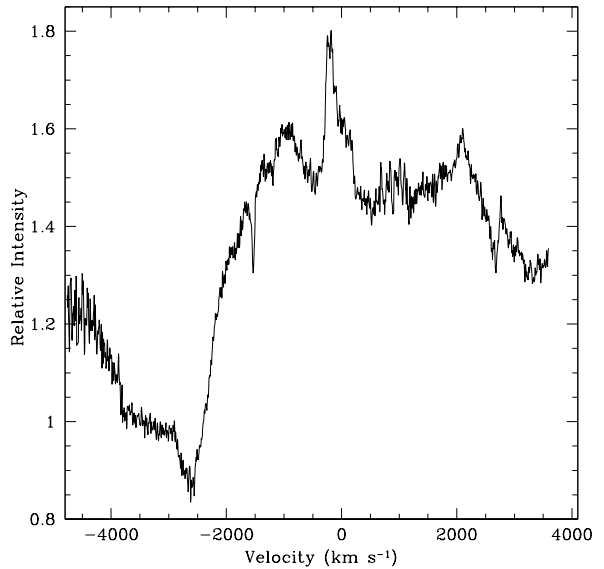}
\vskip -2cm
\caption{The echelle spectrum of the 1.083-$\mu$m He\,{\sc i} line.  Velocities are given with
respect to the line-centre}
\label{1431083.ps}
\end{figure}

The spectra were dereddened assuming A$_V$=5.47 with an interstellar 
reddening law with R$_V$=A$_V$/E(B-V)=3.1.  Line fluxes and equivalent widths 
of the emission lines are estimated from the dereddened spectra.  The 
$J$-band spectra of the three epochs were averaged before estimating
the equivalent widths.  The lines were fitted with Gaussian profiles for
estimating the EWs and FWHM. Multiple Gaussians were fitted when there was 
line blending. For the strongly blended lines, we give only the the 
equivalent widths of the combined profiles;  FWHM of the individual components 
are not listed in those cases. The lines identified, their fluxes and the
equivalent widths are shown in Table {\ref {lineid1}}.  The line 
identifications  are adopted from Eenens, Williams \& Wade (1991) and from 
the atomic line list of Peter van Hoof {\footnote{van Hoof, P. A. M., 
http://star.pst.qub.ac.uk/$^\sim$pvh/}}.  The arc lines observed using the 
UIST high-resolution grisms showed an average FWHM of 75\,km\,s$^{-1}$.  
The FWHM of the narrow emission lines measured in these spectra have been 
corrected to account for this.  The identifications, FWHM, flux and EW of the 
narrow emission lines are listed in Table \ref {lineid2}. The estimates for 
the broad WR emission lines from these high-resolution spectra are listed 
in Table \ref {lineid1}. Most of the errors in the line
fluxes and the EWs arise from the accuracy with which the continuum is defined.
Hence, the errors are determined by multiple measurements of these values on any
specific line or blend. The values given in Table \ref {lineid1} and  \ref {lineid2}
are the averages of eight independent determinations from the dereddened spectra.
The 1$\sigma$ standard deviations are given in brackets against the line 
fluxes and the EWs.

\begin{figure*}
\includegraphics[width=17cm,clip]{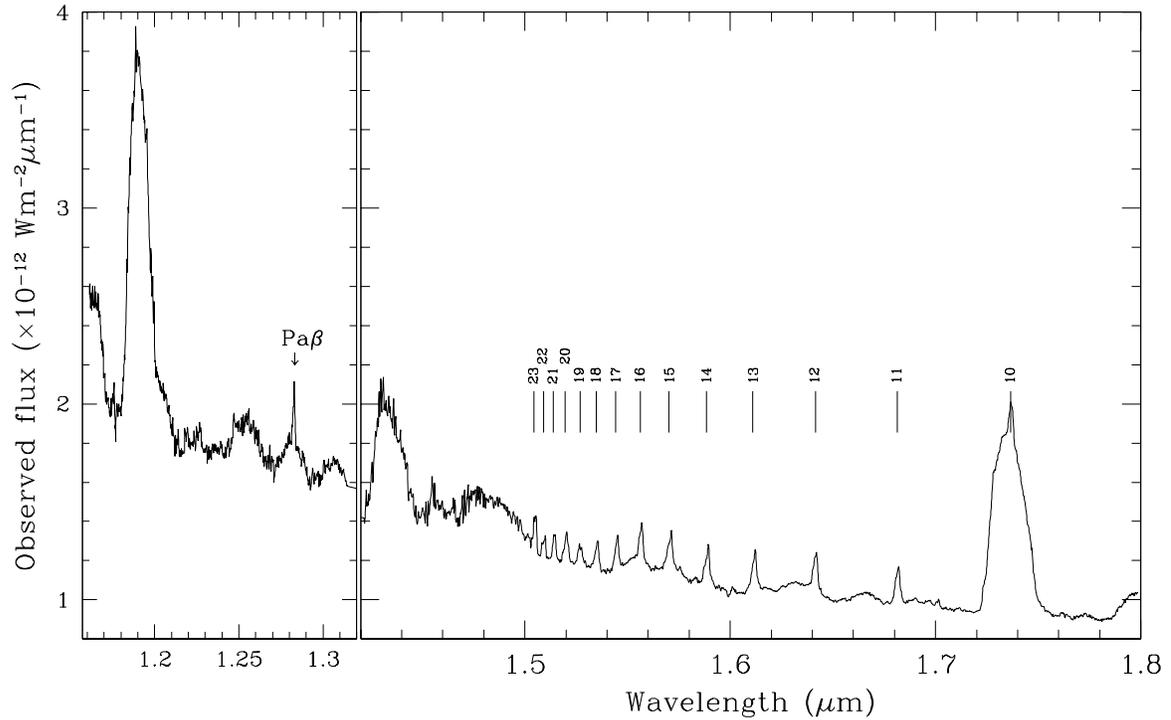}
\vskip -5cm
\caption{High-resolution spectrum of WR 143 observed using UIST. The upper levels
of the H\,{\sc i} Br series lines are indicated against the lines as numbers. }
\label{wr143JH_U}
\end{figure*}

\begin{table*}
\begin{centering}
\caption{Line identifications, Equivalent widths, line fluxes and FWHM 
estimated from the dereddened spectra.  A blank line separates the blends. 1$\sigma$ errors are shown in brackets.
Line strengths estimated from the high-resolution spectra are shows by ``*''. The rest are from the low-resolution spectra.}
\label{lineid1}
\begin{tabular}{@{}llclll}
\hline
$\lambda$ 	&Main contributor &Other possible contributors&Flux 				&EW		& FWHM				\\
($\mu$m)        &		  &			      &($\times$10$^{-15}$Wm$^{-2}$)	&(\AA)		&(km s$^{-1}$) 			\\
\hline
1.054		&C\,{\sc iv} (12--9)    &1.055 (C\,{\sc iii}); 1.055 (O\,{\sc iii})  		&28 (0.8)	&-21 (0.6)   	&2715 (43)    	\\[0.8mm]
1.083 +		&He\,{\sc i} (2p--2s); 	&							&64.3 (0.5)	&-54.5 (0.5)  	&      		\\[-.75mm]
\,\,\,\,1.092 +         &He\,{\sc i} (6f--3d);  &							&		&		&               \\[-.75mm]
\,\,\,\,1.094           &He\,{\sc ii} (12--6)   &							&               &               &               \\[0.8mm]
1.11  		&O\,{\sc iv} (3p--4f)  	&1.1--1.101 (He\,{\sc i})				&11.7 (0.6)	&-12.6 (0.6)  	&3340 (112) 	\\[0.8mm]
1.139   	&C\,{\sc iv} (8d--7p)  	&1.14 (O\,{\sc iii})				        &6.5 (0.4)	&-6.6 (0.4)     &		\\[0.8mm]
1.163 +   	&He\,{\sc ii} (7--5);  	& 							&48 (0.8)	&-53 (1)	&3070 (140)	\\[-.75mm]
\,\,\,\,1.162--1.163 +  &C\,{\sc iv} (14-10);	&						&		&		&		\\[-.75mm]
\,\,\,\,1.164		&C\,{\sc iii} (7d-6p)	                                                &               &               &               \\[0.8mm]
1.191 +  	&C\,{\sc iv} (8--7);    &1.191 (O\,{\sc v});    				&123 (3)	&-145 (4)    	&        	\\[-.75mm]
\,\,\,\,1.198--1.199   	&C\,{\sc iii} (4p--4s) 	&1.197 (He\,{\sc i})					&		&		&	 	\\[-.75mm]
1.205   	&O\,{\sc v} (2p.4d--2p.4p)    &1.20 (O\,{\sc iv}); 1.2--1.209 (O\,{\sc iii}) 	&17 (0.6)	&-20.7 (0.6) 	&2845 (30)   	\\[0.8mm]
1.226   	&C\,{\sc iv} (8p--7d)   &               					&3 (0.5)	&-3.9 (0.7)     &2670 (78)    	\\[0.8mm]
1.255--1.26   	&C\,{\sc iii} (7f--6d, 9-7)  &1.253 (He\,{\sc i}); 1.258, 1.261 (C\,{\sc iii});	&16.6 (0.2)  	&-23.6 (0.3)	&        	\\[-.75mm]
		&                       &1.247, 1.255 (O\,{\sc v}); 1.25 (O\,{\sc iii})		&		&		&		\\[0.8mm]
1.281 +	   	&He\,{\sc ii}(10--6);  	&1.282 (O\,{\sc iv})					&10 (0.2)	&-15.4 (0.3) 	&2530 (25) 	\\[-.75mm]
\,\,\,\,1.279,1.285&He\,{\sc i} (5--3)	&						&		&		&		\\[0.8mm]
1.298, 1.307	&C\,{\sc iv} (8s--7p, 16--11)&1.297, 1.299 (He\,{\sc i});			&7.8 (0.4)	&-13 (0.7)     	&		\\[-.75mm]
		&			&1.3 (O\,{\sc iii}); 1.31 (O\,{\sc iv})   		&       	&		&        	\\[0.8mm]
1.435		&C\,{\sc iv} (4p--4s)   &         						&23 (1)		&-54.7 (2.7)    &2800 (33)    	\\[0.8mm]
1.454		&C\,{\sc iii} (7g--6f)  &                					&1.6 (0.2)	&-4.2 (0.6)     &1950 (123)    	\\[0.8mm]
1.476   	&He\,{\sc ii} (9--6)    &1.474 (O\,{\sc iii}); 1.473 (O\,{\sc v}); 1.473 (C\,{\sc iii}); 1.476 (C\,{\sc iv})	&6.5 (0.2)	&-17.8 (0.5)    &  \\[-.75mm]
1.489 +   	&He\,{\sc ii} (14--7);  &               					&5.6 (0.3)	&-16.1 (0.8)    &        	\\[-.75mm]
\,\,\,\,1.490, 1.491   	&C\,{\sc iv} (15f--11d, 15g--11f)	&				&       	&		&        	\\[0.8mm]
*1.546,1.552   &He\,{\sc i} (11--4, 13--4)    &1.55 (O\,{\sc iv})				&3.8 (0.1)	&-13.6 (0.3)   	&		\\[0.8mm]
*1.570 +   	&He\,{\sc i} (15--4); 	&               					&2.1 (0.1)	&-7.7 (0.2)  	&		\\[-.75mm]
\,\,\,*1.572          &He\,{\sc ii} (13--7)   &                                                 &		&		&               \\[-.75mm]
*1.579, 1.587   &He\,{\sc i} (12-4, 14-4)   &1.58-1.587 (O\,{\sc v})		               	&1.04 (0.2)	&-4 (0.7)     	&		\\[0.8mm]
*1.616, 1.61   	&He\,{\sc i} (11--4, 13-4)  &1.61,1.619 (O\,{\sc v})   		        &0.83 (0.1)	&-3.4 (0.5)     &	   	\\[0.8mm]
*1.623 +   	&C\,{\sc iii} (7p--6d); &1.63-1.64 (He\,{\sc i}); 1.632 (O\,{\sc v})	   	&3.9 (0.2)	&-16.4 (0.6)    &		\\[-.75mm]
\,\,\,\,*1.635          &C\,{\sc iv} (17--12)   &							&		&		&               \\[0.8mm]
*1.664   	&C\,{\sc iv} (9d--8p)   &1.668 (He\,{\sc i})		               		&1.4 (0.1)	&-6.1 (0.4)     &		\\[0.8mm]
*1.693   	&He\,{\sc ii} (12--7) 	&1.681-1.701 (He\,{\sc i}) 				&1.3 (0.2)	&-6.2 (0.8)     &		\\[0.8mm]
1.732-1.740	&C\,{\sc iv} (9--8)     &1.734 (He\,{\sc i})					&40.8 (.5)	&-208 (3)   	&2870 (12)    	\\[0.8mm]
1.790,1.801     &C\,{\sc iv} (9--8, 14--11)&1.81 (C\,{\sc iv}); 1.814 (He\,{\sc i})		&16.8 (0.5)	&-95.7 (2.9)&        	\\[.8mm]
1.944		&He\,{\sc i} (8-4)	&       						&2.0 (0.03)	&-15.2 (.3)    	&1450 (19)    	\\[0.8mm]
2.010		&C\,{\sc iv} (18--13)   &							&1.1 (0.1)	&-8.5 (0.8)     &        	\\[0.8mm]
2.071+   	&C\,{\sc iv} (3d--3p);  &2.061 (He\,{\sc i})					&80.9 (0.3)	&-729 (4) 	&    		\\[-.75mm]
\,\,\,\,2.059	&He\,{\sc i} (2p-2s)	&						&		&               &               \\[-.75mm]
2.108, 2.122+  	&C\,{\sc iii} ( 5p-5s, 4d-4p);	&2.107 (C\,{\sc iv});			 	&13.6 (0.4)	&-129 (4)  	&      		\\[-0.75mm]
\,\,\,\,2.113	&He\,{\sc i} (4s-3p)    &2.123 (O\,{\sc iv})     			&       	&		&        	\\[0.8mm]
2.165		&He\,{\sc i} (7--4)	&               					&2.2 (0.2)	&-22.3 (2)  	&        	\\[-.75mm]
2.189		&He\,{\sc ii} (10--7)	&               					&2.7 (0.2)	&-28.9 (2)   	&        	\\[0.8mm]
2.227		&He\,{\sc i} (7s--4p)	&							&0.79 (0.15)	&-9.04 (1.7)	&		\\[0.8mm]
2.278   	&C\,{\sc iv} (15-12)  	&               					&2.86 (0.1)	&-34 (1)    	&2755 (40)   	\\[0.8mm]
2.318   	&C\,{\sc iv} (17--13)   &2.314 (He\,{\sc ii}); 2.323--2.327 (C\,{\sc iii}); 2.328 (C\,{\sc iv}) &1.7 (0.05)&-21 (0.6)   &        	\\[0.80mm]
2.423--2.427 + 	&C\,{\sc iv} (13--11);  &       						&18.4 (0.3)	&-259 (5)    	&        	\\[-.75mm]
\,\,\,2.422--2.433   	&C\,{\sc iv} (10--9)   	&2.432 (O\,{\sc v})				&      		&		&        	\\[0.8mm]
2.473   	&He\,{\sc i} (6d--4p) 	&2.470 (O\,{\sc v}) 					&4.6 (0.2)	&-74 (4)     	&        	\\[-.75mm]
\hline
\end{tabular}
\end{centering}
\end{table*}

\begin{table}
\begin{centering}
\caption{Narrow emission lines }
\label{lineid2}
\begin{tabular}{@{}lllll}
\hline
$\lambda$ observed &Identification  &Flux$\times$10$^{-16}$	    &EW\footnotemark[1] & FWHM \\
($\mu$m)  &	 &(Wm$^{-2}$)	         &(\AA) &(km\,s$^{-1}$) \\
\hline
\multicolumn{5}{l}{Hydrogen lines}  \\
1.2827		&5-3      &1.5 (0.1)     &-2.2 (0.1)  &306 (7)      \\
1.50495 	&23-4     &		&	    &     	 \\
1.50938 	&22-4     &5.1 (0.1)     &-1.6 (0.04) &389 (4)      \\
1.51438 	&21-4     &6.7 (0.2)     &-2.1 (0.1)  &334 (6)      \\
1.52021 	&20-4     &9.3 (0.3)     &-3.0 (0.1)  &465 (12)     \\
1.52692 	&19-4     &7.0 (0.2)     &-2.3 (0.1)  &511 (11)     \\
1.53517  	&18-4	  &8.0 (0.1)     &-2.7 (0.04) &418 (5)      \\
1.54485 	&17-4     &8.6 (0.2)     &-2.9 (0.1)  &407 (6)      \\
1.55665  	&16-4     &9.9 (0.3)     &-3.3 (0.1)  &395 (11)     \\
1.57082  	&15-4     &13.7 (0.3)    &-4.8 (0.1)  &591 (11)     \\
1.58895 	&14-4     &12.4 (0.5)    &-4.7 (0.2)  &539 (15)     \\
1.61179 	&13-4     &12.6 (0.2)    &-5.1 (0.1)  &515 (5)      \\
1.64153 	&12-4     &12.4 (0.3)    &-5.1 (0.1)  &485 (8)	 \\
1.68164 	&11-4     &11.6 (0.2)    &-5.3 (0.1)  &484 (5)    	 \\
1.73678	&10-4     &8.3(0.4)	&-2.1(0.1)  &339 (14)     \\
2.166   &\multicolumn{4}{l}{7-4  ---- detected in the low-resolution spectrum }\\[1.6mm]
\multicolumn{5}{l}{Helium line} \\
1.083   &2p$^3$P-2s$^3$S &3.6 (0.1)     &-2.4 (0.1)  &          \\[1.6mm]
\hline
\multicolumn{5}{l}{\footnotemark[1] Will be strongly affected by the Wolf-Rayet continuum and the }\\
\multicolumn{5}{l}{broad emission lines}\\

\end{tabular}
\end{centering}
\end{table}

\section{Discussion and Conclusions}

Table \ref{mags} and Fig. \ref{wr143flux.ps} show the variability of the
$JHK$ magnitudes.
The $JHK$ magnitudes measured by us on the three epochs (within $\sim$21
months) did not exhibit any significant variability beyond the observational
errors.  However these three measurements are consistently brighter than
the 2MASS $JHK$ magnitudes observed 3 years prior to our first measurement
by $\sim$0.45 magnitudes.  To verify if this is genuine, we measured the 
$JHK$ magnitudes of two other stars in our images and compared those with 
the 2MASS measurements and found a good agreement. Hence, the brightening 
that we observe in our photometry compared to 2MASS could indicate a
genuine brightening of one of the stars in the WR 143 binary.

From the dereddened spectra, we estimated the ratios of the equivalent 
widths of the emission lines
(1.083+1.094)/(1.191+1.199) and (1.693+1.701)/1.736 to be 0.38 and
0.03 respectively which are consistent with a WC type earlier than WC5 when 
compared with the line ratios estimated by Eenens et al. (1991).
The ratios  1.28/(1.083+1.094), 2.08/2.11 and 2.43/2.48  
are 0.28, 5.7 and 3.5 respectively, which are somewhat less for
WC4 star.  However, the line ratios estimated by Eenens et al. (1991)
extend down only up to WC5 type and have only one or two objects per
WC type, and hence we do not know about the uncertainty in these ratios.
In general our
$JHK$ spectra agree with a WC4 type for this star.

WR 143 was detected by the MSX at 8.28\,$\mu$m (Egan et al. 2003) with a 
flux density of 0.1285\,mJy. Observations
at 10\,$\mu$m by Cohen et al. (1975) and Smith and Houck (2001)  differ very
much and are only upper limits. Hence those are not considered here.
The near-IR magnitudes given in Table \ref{mags} and the line free
magnitudes reported by Massey (1984) (13.97, 13.16, 11.95 and 11.17 magnitudes
respectively in the narrow-band u,b,$\upsilon$ and r filters) are dereddened 
adopting the value of A$_V$=5.47 and following
the extinction relations given by Cardelli, Clayton \& Mathis (1989).
Fig. {\ref{wr143flux.ps} shows the flux density distribution of WR 143 from
0.365\,$\mu$m to 8.28\,$\mu$m.
At radio and infrared wavelengths,
flux density distribution (S$_{\nu}$) of a uniform, ionized, spherically 
symmetric wind can be represented by a power law of the form
S$_{\nu} \propto \nu^{\alpha}$ with $\alpha$=0.6, where ${\nu}$ is the
frequency of observation (Wright \& Barlow 1975).  
This value of $\alpha$ is intermediate between 
$\alpha$=-0.1, expected for free-free emission from 
an optically thin homogeneous plasma and $\alpha$=2.0, expected for that
from an optically thick plasma.
Williams (1999) estimated an average value of  $\alpha$=0.7 from millimeter
and radio observations.
The observed values of $\alpha$ at shorter wavelengths are somewhat higher
since the radiation at these wavelengths are mainly emitted from the inner 
regions where the wind is still being accelerated.
Morris et al. (1993) found that in the 
wavelength interval $\sim$0.14-1.0 $\mu$m,  the continuum energy distribution 
can be represented by a power law with mean $\alpha$=0.85$\pm$0.26.
Setia Gunawan (2001) estimates an average value of 
$\alpha$=1.21$\pm$0.24 from the optical and near-IR (also MSX, mm and radio 
data for some objects) of 9 non-dusty WR stars. Average of the $\alpha$
for six single WR stars given in Setia Gunawan (2001) gives 1.13$\pm$0.24.
A linear least square fit to all the observed data of WR 143 from 0.36\,$\mu$m
to 8.28\,$\mu$m (the continuous line in Fig. {\ref{wr143flux.ps}) gives 
$\alpha$=1.73 (1$\sigma$=0.06). In Fig. \ref{wr143flux.ps},  we have also 
plotted the power law flux density distribution for $\alpha$=1.13 
(the dotted line), scaled to match our fitted line at 1.25\,$\mu$m.
We see that, for WR 143,  the flux density distribution is much steeper
than what is seen for most of the WR stars.  
We propose that the much steeper slope of the SED of WR 143
is due to the increasing contribution of the early B-type
companion star towards shorter wavelengths.  A linear fit, when forced through
the 0.36--3.6 $\mu$m region of a B0V star gives $\alpha$=1.77.

\begin{figure}
\centering
\includegraphics[width=8.0cm,clip]{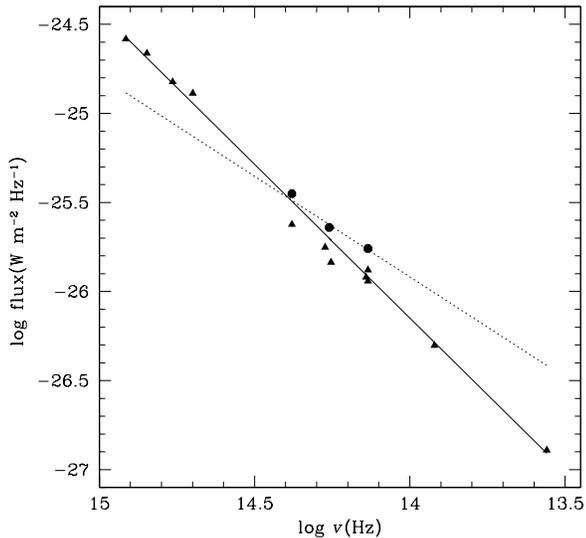}
\vskip -2.5cm
\caption{Flux density distribution of WR 143.
Filled circles show our $JHK$ measurements with the measurements on different
epochs averaged and the filled triangles show observations by others.
The continuous line shows a linear least square fit to the data
which gives $\alpha=$1.72 and the dotted line
is the distribution with $\alpha=$1.13, the average for 6 single WR
stars from the list of Setia Gunawan (2001),
scaled to match our fitted line at 1.25\,$\mu$m.}
\label{wr143flux.ps}
\end{figure}

The presence of the narrow emission lines in our spectra give us additional
clues about the nature of the companion star of WR 143. Most Be
stars in their emission phase show Br series lines in the $H$ and $K$ bands
(Steele and Clark, 2001; Clark and Steele, 2000).
Be stars with spectral type earlier than B2.5 also exhibit the He\,{\sc{i}} line
at 2.058\,$\mu$m in emission (Clark and Steele, 2000).  For WR 143,
this wavelength is blended with the broad C\,{\sc{iv}} and  C\,{\sc{iii}}
emission lines of the WR star; therefore we cannot conclude about its presence 
in the spectrum of the companion.
However, the narrow emission component on the broad 1.083-$\mu$m He\,{\sc{i}}
emission line close to the line centre is stable over the 
epochs of our observation.  This feature is  clearly seen in all the
published near-IR spectra of WR 143 and is identified to be the He\,{\sc{i}}
emission line from the companion. Prominent 1.083-$\mu$m emission line
has been observed in the early-type Be stars before (Lowe et al. 1985).
H\,{\sc{i}} Brackett series emission lines are exhibited by most of the
Be stars in their emission phase.  However the additional presence of the
He\,{\sc{i}} emission line shows that the companion is a Be star of spectral
type earlier than B2.5.

The spectrum of WR 143 appears to undergo long term variations. The 
equivalent width of the 1.083-$\mu$m He\,{\sc i} line measured from our
spectra is 54.5$\pm$0.5\,\AA, which differs from 61\,\AA \, measured by 
Kuhi (1968) and  32\,\AA \, observed in Sept. 1989 (Vreux, Andrillat 
\& Biemont 1990).  The optical spectrum of WR 143 was observed by Torres 
and Massey (1987).  We obtained their published spectrum from The CDS 
Service for Astronomical Catalogues and examined it.  The 
He\,{\sc ii}, C\,{\sc iv} blend at 6564\,\AA \, appears to have a 
sharper peak compared to the other optical emission lines of WR143, 
indicating the possible prsence of H$\alpha$ in emission superposed on 
this blend.  However, it should be noticed that we did not see the
H$\beta$ and H$\gamma$ lines in their spectrum.
The equivalent widths of most of the emission lines are much weaker
in the $K$-band spectrum observed by us compared to those observed by
Figer et al. (1997). We cannot say how much of these variations mentioned
above are due to the differences in the measurement techniques. However, 
from the difference in the equivalent widths between ours and the recent 
measurements of Figer et al. (1997), where both used the same technique,
we can consider some to be genuine.  For the C\,{\sc iv} blends
at 2.071\,$\mu$m and 2.278\,$\mu$m, we measure EW s of 729\,\AA\, and
34\,\AA\, respectively, whereas Figer et al. obtained 1069\,\AA\, and
59\,\AA\, respectively.  To understand if these could be produced by
the brightening that we see in our photometry, we assumed a disk of
T$_{eff}$=18000 K, appropriate for a disk around an early-type Be star
(Waters 1986), as the cause of the current brightening in the 
near IR photometry.  A 18000 K blackbody curve was subtracted from 
our flux-calibrated $K$-band spectrum.  The flux of the blackbody was scaled
so that, when subtracted, the flux of the subtracted spectrum agreed with
that of the observed spectrum calibrated with the 2MASS $K$-band phtometry 
at the band-centre.  The equivalent widths of the lines at
2.071\,$\mu$m and 2.278\,$\mu$m, measured from the subtracted
spectrum, gave 1012\,\AA\, and 52\,\AA\, respectively showing that
the lower EW measured from our spectra compared to that
reported by Figer et al. (1997) is most probably due to a
brightening of the Be star comparison star resulting in an increase in the
continuum flux. Be stars are known to exhibit variability 
(Dougherty and Taylor 1994).

Other than early-type Be stars, a variety of early-type objects like 
Ofpe/WN9 stars and LBV's exhibit H\,{\sc{i}} Pa and Br series emission
lines and He\,{\sc{i}} lines (Morris et al. 1996, Bohannan and
Crowther 1999, Miroshnichenko et al. 2002). However, these
objects are much more luminous than the M$_{\upsilon}$=-3.66 for the
companion derived by van der Hucht (2001) which is typical of 
early-type B and Be stars (Wegner 2000). Hence, we conclude that the 
companion is an early-type Be star.  Since Be Stars are known to vary,
the photometric variations observed must be due to the variability
of the Be star companion.

\section*{Acknowledgments}
UKIRT is operated by the Joint Astronomy Centre, Hilo,
on behalf of the U.K. Particle Physics and Astronomy  Research Council.
Mt. Abu Observatory is operated by the Physical Research Laboratory,
Ahmedabad, funded by the Dept. of Space, Govt. of India.
We would like to thank the UKIRT service program for obtaining some
of the spectra and the staff members of UKIRT and Mt. Abu observatories for helping
us with the data collection. This publication makes use of data products
from the Two Micron All Sky Survey, which is a joint project of the University
of Massachusetts and the Infrared Processing and Analysis Center/California
Institute of Technology.
We thank Peredur Williams for his comments and suggestions, which have improved 
the paper.
Several of the software packages developed by the
Starlink project run by CCLRC on behalf of PPARC and IRAF developed at
NOAO are used for reducing the data. 
This research has also made use of the SIMBAD database
operated by CDS, Strasbourg, France and the NASA ADS database, hosted by the
Harvard-Smithsonian Centre for Astrophysics.

\bsp

\label{lastpage}

\end{document}